\documentclass[epj]{svjour}
\usepackage{graphicx}
\usepackage{amssymb}

\usepackage{subfigure}

\usepackage{setspace}

\def\plotdir{plots}
\newcommand{\DO}{D\O}
\newcommand{\zee}{\mbox{$Z/\gamma^* \rightarrow e^+e^-$}}
\newcommand{\zmumu}{\mbox{$Z/\gamma^* \rightarrow \mu^+\mu^-$}}
\newcommand{\mass}{\mbox{$Q$}}
\newcommand{\dphi}{\mbox{$\Delta\phi$}}
\newcommand{\Qt}{\mbox{$Q_T$}}
\newcommand{\at}{\mbox{$a_T$}}
\newcommand{\al}{\mbox{$a_L$}}

\newcommand{\atom}{\mbox{$a_T/\mass$}}
\newcommand{\matom}{\mbox{$M_Za_T/\mass$}}
\newcommand{\Qtom}{\mbox{$Q_T/\mass$}}
\newcommand{\alom}{\mbox{$a_L/\mass$}}
\newcommand{\atolnm}{\mbox{$a_T/\ln(\mass/Q_0)$}}
\newcommand{\matolnm}{\mbox{$\ln(M_Z/Q_0)a_T/\ln(\mass/Q_0)$}}
\newcommand{\gtwo}{\mbox{$g_2$}}
\newcommand{\ptone}{\mbox{${\vec{p}_T}$$^{(1)}$}}
\newcommand{\pttwo}{\mbox{${\vec{p}_T}$$^{(2)}$}} 

\newcommand{\chisqr}{\mbox{$\chi^{2}$}}
\newcommand{\sts}{\mbox{$\sin(\theta^{*})$}}

\newcommand{\phiaco}{\mbox{$\phi_{\rm acop}$}}
\newcommand{\tanphiaco}{\mbox{$\tan(\phi_{\rm acop}/2)$}}
\newcommand{\phistarr}{\mbox{$\phi^{*}$}}
\newcommand{\phistarCS}{\mbox{$\phi_{\rm CS}^{*}$}}
\newcommand{\phistarEta}{\mbox{$\phi_{\rm \eta}^{*}$}}

\newcommand{\etal}{et al}

\begin{document}
%
\title{\hfill {\small MAN/HEP/2010/12}\\ Optimisation of variables for studying dilepton transverse
  momentum distributions at hadron
  colliders\ \ \ \ \ \ \ \ \ \ }
\author{A. Banfi\inst{1}$^,$\inst{2}
\and S. Redford\inst{3} 
\and M. Vesterinen\inst{3}
\and P. Waller\inst{3}
\and T. R. Wyatt.\inst{3}
}                     
%
%
\institute{ Institute for Theoretical Physics, ETH Zurich, 8093 Zurich, Switzerland.
\and Particle Physics Group, School of Physics and Astronomy, University of Manchester, UK.}
\titlerunning{Optimisation of variables for studying dilepton
  transverse momentum distributions \ldots}
\authorrunning{A. Banfi et al.}
\date{Received: 8/9/2010}
%
\abstract{
In future measurements of the dilepton ($Z/\gamma^*$) transverse momentum, \Qt, at both
  the Tevatron and LHC, the achievable bin widths and the ultimate
  precision of the measurements will be limited by
  experimental resolution rather than by the available event
  statistics.
In a recent paper the variable \at, which corresponds to the component
of \Qt\ that is transverse to the dilepton thrust axis, has been
studied in this regard.
  In the region, \Qt\ $<$ 30 GeV, 
  \at\ has been shown to be less susceptible to experimental resolution and efficiency
  effects than the \Qt.
  Extending over all \Qt, we now demonstrate that dividing \at\ (or \Qt) by the measured dilepton invariant mass
  further improves the resolution.
  In addition, we propose a new variable, \phistarEta, that is determined exclusively
  from the measured lepton directions; this is even more precisely
  determined experimentally than the above variables
  and is similarly sensitive to the \Qt.
  The greater precision achievable using such variables
  will enable more stringent tests of QCD 
  and tighter constraints on Monte Carlo event generator tunes.
} 
\maketitle
%

\section{Introduction}    
\label{sec-intro}
  The production of \zee\ and \zmumu\ at hadron colliders 
  provides an ideal testing ground for the predictions
  of QCD, due to the colourless and relatively background free final state.
  The dilepton transverse momentum, \Qt, distribution
  probes QCD radiation in the initial state.
  At high values of \Qt\ ($\Qt>30$ GeV, say), the fixed order perturbative calculations now available at
  NNLO~\cite{FEWZ,DYNNLO} are expected to yield accurate predictions.
  At low \Qt, soft gluon resummation techniques are required~\cite{CSS1985},
  with additional non-perturbative form factors determined in global fits to experimental data
  such as in~\cite{BLNY2003}.
  Various event generators are also
  available~\cite{Pythia,Herwig,Alpgen,Sherpa}, matching tree level 
  matrix elements to parton showers  tuned to data.
  Validation and tuning (form factors and parton showers) of these models
  require comparison with experimental data that have been corrected
  for detector resolution and efficiency effects.
  Improved understanding of these production models will increase
  sensitivity to new physics signals at hadron colliders.
  
  The \Qt\ distribution has been measured at the Fermilab Tevatron, 
  by the CDF~\cite{CDFRunI} and \DO~\cite{DzeroRunI,DzeroRunII,DzeroDimuZpT} Collaborations.
  The most recent of the above measurements~\cite{DzeroRunII,DzeroDimuZpT} used approximately 1~fb$^{-1}$ of data.
  Although this represents only about one tenth of the anticipated
  final Tevatron data set, the precision of these measurements 
  was already limited by experimental systematic uncertainties in the
  corrections for event selection efficiencies and unfolding of lepton
  momentum resolution.
  In order to unfold measured distributions for experimental
  resolution it is important that the chosen bin widths are not too
  small compared to the experimental resolution.
  In the low \Qt\ region in~\cite{DzeroRunII,DzeroDimuZpT}, 
  the minimum bin sizes were determined  by experimental resolution
  rather than the available data statistics.
  The final Tevatron data set will be an order of magnitude
  larger than that analysed in~\cite{DzeroRunII,DzeroDimuZpT}.
New ideas are therefore needed in order to exploit fully the data
  for studying the physics of boson \Qt.

\begin{figure}[htbp]
    \includegraphics[width=0.45\textwidth]{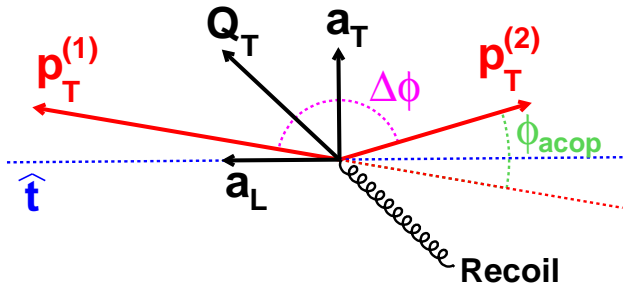}\\
    \caption{Graphical illustration in the plane transverse
  to the beam direction of the variables defined in the
      text and used to analyse dilepton transverse momentum
      distributions in hadron colliders.}
    \label{Figure:at}
  \end{figure}

The \at\ variable was introduced in~\cite{OPAL-llvv} and was proposed
  as a novel variable for studying the \Qt\ in~\cite{Mika_Terry_NIM}.
Figure~\ref{Figure:at} illustrates this and other relevant variables
defined below.
  Events with $\dphi > \pi/2$, where \dphi\ is the azimuthal opening angle of the lepton pair,
  correspond to approximately 99\%  of the total cross section.
  For such events the \Qt\ is split into two components 
  with respect to an event axis defined as, 
$\vec{\hat{t}} = \left(\ptone - \pttwo\right)/|\ptone - \pttwo |$, where
  \ptone\ and \pttwo\ are the lepton momentum vectors in the plane transverse
  to the beam direction.
  The component transverse to the event axis is denoted by \at\ and the
  aligned component is denoted by \al.
  For events with $\dphi < \pi/2$ this decomposition is not useful and
  \at\ and \al\ are defined as being equal to \Qt.
  \at\ is less susceptible than \Qt\ to the lepton $p_T$ resolution.
  In addition, the efficiencies of selection cuts on lepton isolation and $p_T$ 
  are shown in~\cite{Mika_Terry_NIM} to be less correlated with  \at\ than \Qt.
  For studying the low \Qt\ (non-perturbative) region, 
  \at\ is thus a more powerful variable than \Qt.
  The \at\ distribution has subsequently been calculated  to NLL accuracy
  using soft gluon resummation techniques~\cite{Banfi_aT}.

  A recent paper~\cite{dphi} has discussed the idea of using 
  \dphi, as an analysing  variable that is sensitive to the physics of \Qt,
  and insusceptible to lepton momentum uncertainties\footnote{We note
    that the expected
    distribution of \dphi\ does have a small residual
    sensitivity to the lepton $p_T$ measurement.
This arises from
    the cut on \mass\ in the event sample
    selection, which is affected by the lepton $p_T$ scale and resolution.}. 
  Whilst \dphi\ is primarily sensitive to the same component of \Qt\ as \at,
  the translation from \at\ to \dphi\ depends on the scattering angle
  $\theta^*$ of the leptons relative to the beam direction in the dilepton rest frame. 
  Thus, \dphi\ is less directly related to \Qt\ than \at. 
  As a result, \dphi\ has somewhat smaller statistical sensitivity 
  to the underlying physics than \at.
  
  In this paper, we discuss two further ideas to improve experimental precision,
  whilst maintaining (\Qt) physics sensitivity:
  \begin{itemize}
  \item Dividing \at, and \Qt\ by the dilepton invariant mass, \mass,
    thus further reducing the effects of lepton $p_T$ resolution, and 
    almost totally cancelling lepton $p_T$ scale calibration uncertainties.
  \item Correcting \dphi\ on an event-by-event basis for the
    scattering angle, $\theta^*$, thus improving the sensitivity to \Qt.
  \end{itemize}

An overview of the rest of this paper is as follows.
In Sect.~\ref{sec-massratio} we give an approximate analytic
motivation for the idea of dividing \at\ (and \Qt) by \mass\ in order
to produce variables with very substantially improved experimental resolution.
In Sect.~\ref{sec-costheta} we discuss the idea of correcting \dphi\ on an event-by-event basis for the
    scattering angle, $\theta^*$, thus improving the sensitivity to
    \Qt. In addition, we propose a new variable,
    $\cos(\theta^{*}_{\eta})$, which provides a measure of the scattering angle that
  is based entirely on the measured track directions and is thus
  extremely well measured experimentally.
In Sect.~\ref{fast_sim} we describe the simple parameterised detector
simulation we employ in our MC studies.
In Sect.~\ref{scaling_factors}--\ref{sensitivity} we present the
results of our MC studies of the performance of the various candidate
variables in terms of their experimental resolution and their
sensitivity to the underlying physics.
In Sect.~\ref{conclusions} we present some conclusions, including our recommendations for the best
variables to use in experimental studies of the transverse momentum of
dilepton pairs produced at hadron colliders.

  

\section{Mass ratios of \boldmath{\at} and \boldmath{\Qt}}
\label{sec-massratio}
  For $\dphi \approx \pi$, \at\ is given by the approximate formula
  $$ \at = 2 \frac{p_T^{(1)}p_T^{(2)}}{p_T^{(1)}+p_T^{(2)}}\sin\dphi$$ 
  and thus the fractional change in \at\ with respect to a variation in,
  say, 
  $p_T^{(1)}$ is given by 
  $$\frac{\Delta \at}{\at} = \frac{p_T^{(2)}}{p_T^{(1)}+p_T^{(2)}}\frac{\Delta p_T^{(1)}}{p_T^{(1)}}. $$
  The dilepton invariant mass is given by
  $$\mass = \sqrt{2p^{(1)}p^{(2)}(1-\cos\Delta\theta)},$$
  where
  $p^{(1)}$ and $p^{(2)}$ are the lepton momenta and $\Delta\theta$ is
  the angle between the two leptons.
  Thus, the fractional change in mass with respect to a variation in $p^{(1)}$ is given by 
  $$\frac{\Delta \mass}{\mass} = \frac{1}{2}\frac{\Delta p^{(1)}}{p^{(1)}}. $$
  Since track angles are extremely well measured it can be taken to a
  very good approximation that
  $$\frac{\Delta p_T^{(1)}}{p_T^{(1)}}= \frac{\Delta p^{(1)}}{p^{(1)}}.$$
  The fractional change in \atom\ with respect to a variation in
  $p^{(1)}$ is thus given by 
  $$\frac{\Delta \left(\at/\mass\right)}{\left(\at/\mass\right)} 
  = \frac{\Delta \at}{\at} - \frac{\Delta \mass}{\mass} 
  = \left(\frac{p_T^{(2)}}{p_T^{(1)}+p_T^{(2)}} - \frac{1}{2}\right)
  \frac{\Delta p_T^{(1)}}{p_T^{(1)}}. $$
  
  Thus the variations with $p_T^{(1)}$ in \at\ and $\mass$ partially cancel in the ratio,
  rendering \atom\ less susceptible to the effects of lepton $p_T$ resolution than \at.
  In particular, in the region of low \Qt\ then $p_T^{(1)} \approx p_T^{(2)}$ and thus 
  $\Delta (\at/\mass) \approx 0$.
  Similarly, the quantity $\Qt/\mass$ is  less susceptible to the effects of lepton $p_T$ resolution than \Qt.
  
  A simple example of an uncertainty in the lepton $p_T$ scale
  calibration is to consider the $p_T$ of all leptons to be multiplied
  by a constant factor.
  It can be seen trivially that in this case \at, \Qt\ and \mass\ are
  all multiplied by the same factor and that the measured \atom\ and \Qtom\ are
  unaffected by such a scale uncertainty.
  
  
  
\section{Correcting \boldmath{\dphi} for the scattering angle}
\label{sec-costheta}

  The azimuthal opening angle between the two leptons, \dphi,
  is primarily sensitive to the same component of \Qt\ as \at, and is based only on the well
  measured lepton angles.
  However, at fixed $\atom$, \dphi\ depends on the scattering angle
  $\theta^*$ of the leptons relative to the beam direction in the dilepton rest frame. 
  For convenience, we define the acoplanarity angle, \phiaco, as $\phiaco = \pi-\dphi$.
  For $p_T^{(1)} \approx p_T^{(2)}$ it can be fairly easily shown
  that 
$$\atom \approx \tan(\phiaco/2)\sin(\theta^{*}).$$
  This suggests that the variable 
$$\phistarr \equiv \tan(\phiaco/2)\sin(\theta^{*})$$ 
may be an appropriate alternative
  quantity with which to study \Qt.
  
  
  In the analysis of hadron-hadron collisions,  $\theta^{*}$ is
  commonly evaluated in the Collins Soper frame~\cite{CS}.
  However, $\theta^{*}_{\rm CS}$ requires knowledge of the lepton
  momenta and is thus susceptible to the effects of lepton momentum
  resolution.
  Motivated by the desire to obtain a measure of the scattering angle that
  is based entirely on the measured track directions (since this will
  give the best
  experimental resolution) we propose here an alternative definition of
  $\theta^{*}$.
  We apply a Lorentz boost along the beam direction such that the two
  leptons are back-to-back in the $r$-$\theta$ plane.
  This Lorentz boost corresponds to $\beta=\tanh\left(\frac{\eta^-+\eta^+}{2}\right)$
  and yields the result\footnote{The 
lepton pseudorapidity, $\eta$, is defined as $\eta = -\ln[\tan(\frac{\theta}{2})]$, where $\theta$ is 
  the polar angle with respect to the beam direction, in the laboratory frame.}
  $$
  \cos(\theta^{*}_{\eta})=\tanh\left(\frac{\eta^--\eta^+}{2}\right), 
  $$
where $\eta^-$ and $\eta^+$ are the pseudorapidities of the negatively
and positively charge lepton, respectively.

  We  consider two candidate variables
  $$\phistarCS  \equiv \tanphiaco\sin(\theta^{*}_{\rm CS})$$ 
  $$\phistarEta \equiv \tanphiaco\sin(\theta^{*}_{\eta})$$
  for further evaluation  in terms of their experimental resolution
  and physics sensitivity.


  \section{Simple parameterised detector simulation}
  \label{fast_sim}
  Monte Carlo events are generated using {\sc pythia}~\cite{PYTHIA},
  for the process $p\bar{p} \rightarrow Z/\gamma^*$, in the $e^+e^-$ and $\mu^+\mu^-$ decay channels,
  and re-weighted in dilepton \Qt\ and rapidity, $y$, to match the higher order predictions
  of {\sc resbos}~\cite{ResBos} as in~\cite{Mika_Terry_NIM}.
  Electrons and muons are defined at ``particle level'' according to the prescription in~\cite{LH09},
  and at ``detector level'' by applying simple Gaussian smearing to the particle level momenta:
  $\delta(1/p_T) = 3\times10^{-3}$ (1/GeV) for muons, which are
  measured in the tracking detectors; $\delta p/p = 0.4(p/p_0)^{-1/2}$
  with $p_0 = 1$~GeV for
  electrons, which are measured in the calorimeter.
  In addition, the particle angles are smeared, assuming Gaussian resolutions of $0.3\times10^{-3}$~rad for $\phi$ and $1.4\times10^{-3}$ for $\eta$.
  These energy, momentum, and angular resolutions roughly correspond to those in the 
  \DO\ detector~\cite{DZERO-TDR}.
  
  Events are accepted for further analysis if: $70 < \mass < 110$~GeV and both leptons satisfy the
  requirements $p_T > 15$~GeV and $|\eta|<2$.
  These selection cuts are made at particle level, unless otherwise stated.


\section{Scaling factors}
  \label{scaling_factors}

In the following sections, we compare the experimental resolution and physics sensitivity 
of the various candidate variables.
In particular, we compare the variation of the resolution for each variable as a function of that variable.
This comparison is facilitated by ensuring that all distributions have approximately the same scale and shape.
Compared to \Qt\ or \Qtom, all other variables are on average a factor
$\sqrt{2}$ smaller (since \at\ and \al\ measure one component of \Qt).
A simple multiplication by $M_Z$ (= 91.19 GeV~\cite{PDG}) corrects for the average $Q^{-1}$ factor
in the mass ratio and angular variables and conveniently ensures that all variables have the same 
units (GeV).
Finally, 
the mean value of $\sin(\theta^*)$ is around $\sim 0.85$, and \tanphiaco\ is scaled by this
additional factor.
The above factors are summarised in Table~\ref{Table:ScalingFactors}.

  \begin{table}[ht]
    \centering
    \begin{tabular}{c c }
      \hline\hline  
      variable & scaling factor \\
      \hline
      \Qt         &  1 \\
      \Qtom       &  $M_Z$ \\
      \at         &  $\sqrt{2}$ \\
      \atom       &  $\sqrt{2}M_Z$ \\
      \al         &  $\sqrt{2}$ \\
      \alom       &  $\sqrt{2}M_Z$ \\
      \tanphiaco  &  $0.85\sqrt{2}M_Z$ \\
      \phistarCS  &  $\sqrt{2}M_Z$ \\
      \phistarEta &  $\sqrt{2}M_Z$ \\
      \hline\hline

    \end{tabular}
    \caption{Scaling factors for different candidate variables.}
    \label{Table:ScalingFactors}
  \end{table}

  
  \section{Experimental resolution for dilepton scattering angle}
  \label{sec-resolution_costheta}
  
\begin{figure}[htbp]
    \subfigure[$70 < Q < 110$ GeV]{\includegraphics[width=0.37\textwidth]{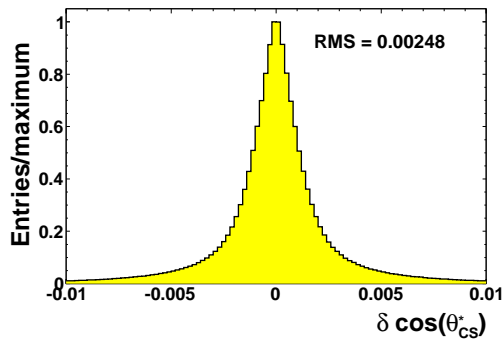}}
    \subfigure[$70 < Q < 110$ GeV]{\includegraphics[width=0.37\textwidth]{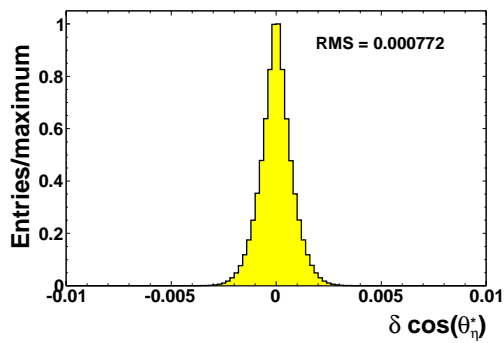}}
    \subfigure[$500 < Q < 600$ GeV]{\includegraphics[width=0.37\textwidth]{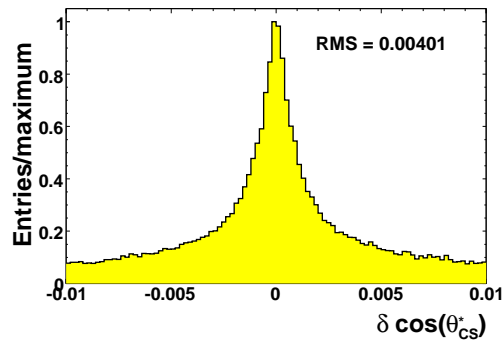}}
    \subfigure[$500 < Q < 600$ GeV]{\includegraphics[width=0.37\textwidth]{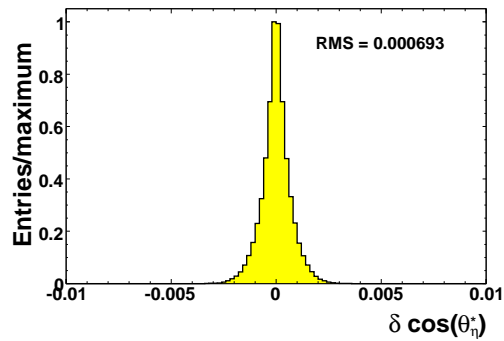}}
    \caption{Distributions of experimental resolutions the dilepton centre of mass scattering
    angle for events satisfying  $70 < Q < 110$~GeV (upper row) and  $500 < Q < 600$~GeV (lower row).
      Figures (a) and (c) show $\cos(\theta^{*}_{\rm CS})$. Figures (b) and (d) show $\cos(\theta^{*}_{\eta})$.}
    \label{Figure:resolution_costheta}
  \end{figure}

  Figure~\ref{Figure:resolution_costheta} shows the experimental
  resolution of $\cos(\theta^{*}_{\rm CS})$ and   
$\cos(\theta^{*}_{\eta})$ in our simulation of dimuon events.
  The upper row of Figure~\ref{Figure:resolution_costheta} shows events
  that satisfy $70 < Q < 110$~GeV; it demonstrates that
  $\cos(\theta^{*}_{\eta})$ is significantly
  better measured experimentally than  $\cos(\theta^{*}_{\rm CS})$.
  This is because $\cos(\theta^{*}_{\eta})$ is evaluated using only angular measurements, which
  are more precise than the momentum measurements 
  included in the determination of $\cos(\theta^{*}_{\rm CS})$.

The variable $\cos(\theta^{*}_{\eta})$ is used in the definition
of $\phistarEta = \tanphiaco\sin(\theta^{*}_{\eta})$ in Section~\ref{sec-costheta} above.
As an aside, we note in addition that a precise determination of the dilepton centre of mass scattering
    angle that is free from the effects of lepton momentum resolution can also find application in the determination of the
    forward-backward charge asymmetry of dilepton production at
    hadron colliders.
The experimental resolution in $\cos(\theta^{*}_{\rm CS})$ becomes particularly
significant in the dimuon channel for very high values of $Q$ for
which the lepton momentum resolution is poorest.
This is illustrated in  the lower row of
Figure~\ref{Figure:resolution_costheta}, which shows the experimental
  resolution of $\cos(\theta^{*}_{\rm CS})$ and   
$\cos(\theta^{*}_{\eta})$ in events
  that satisfy  $500 < Q < 600$~GeV.
The advantage of using  $\cos(\theta^{*}_{\eta})$ for high mass events
is even larger than
was the case for $70 < Q < 110$~GeV.

  \section{Experimental resolution for variables related to  dilepton \boldmath{\Qt}}
  \label{sec-resolution}
  
\begin{figure}[htbp]
    \centering
    \subfigure[Tracker]{\includegraphics[width=0.48\textwidth]{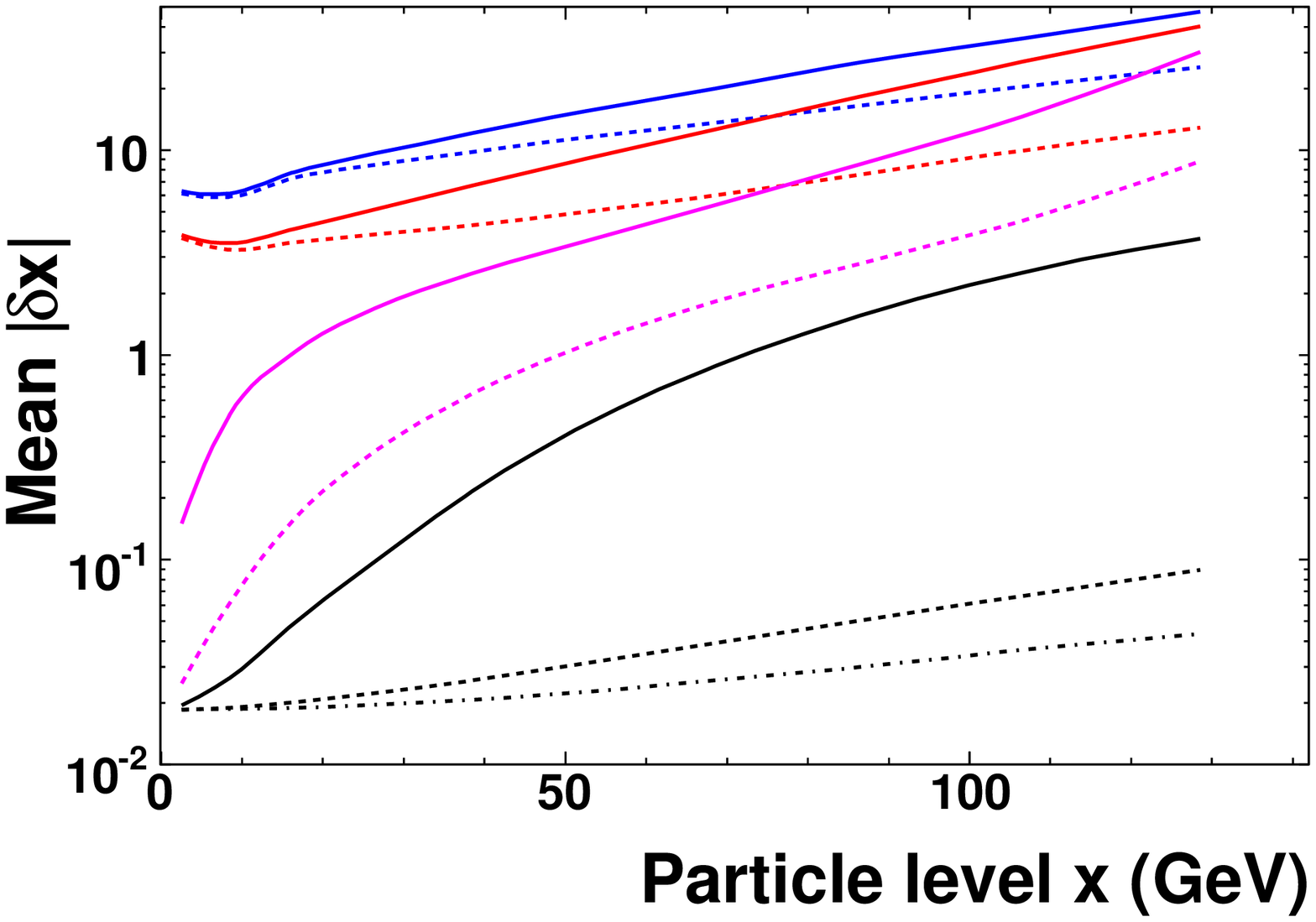}}\\
    \subfigure[Calorimeter]{\includegraphics[width=0.48\textwidth]{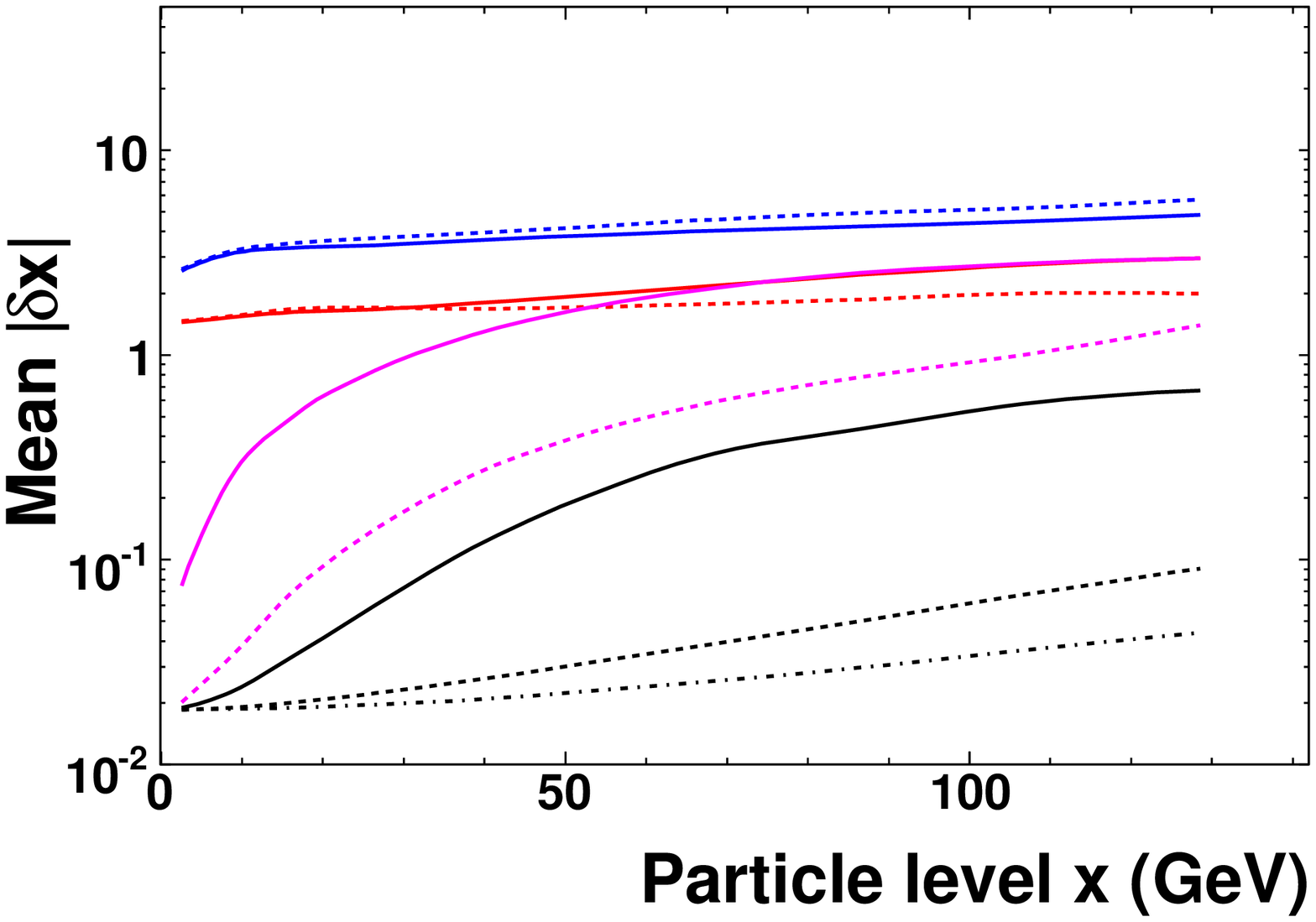}}
    {\includegraphics[width=0.45\textwidth]{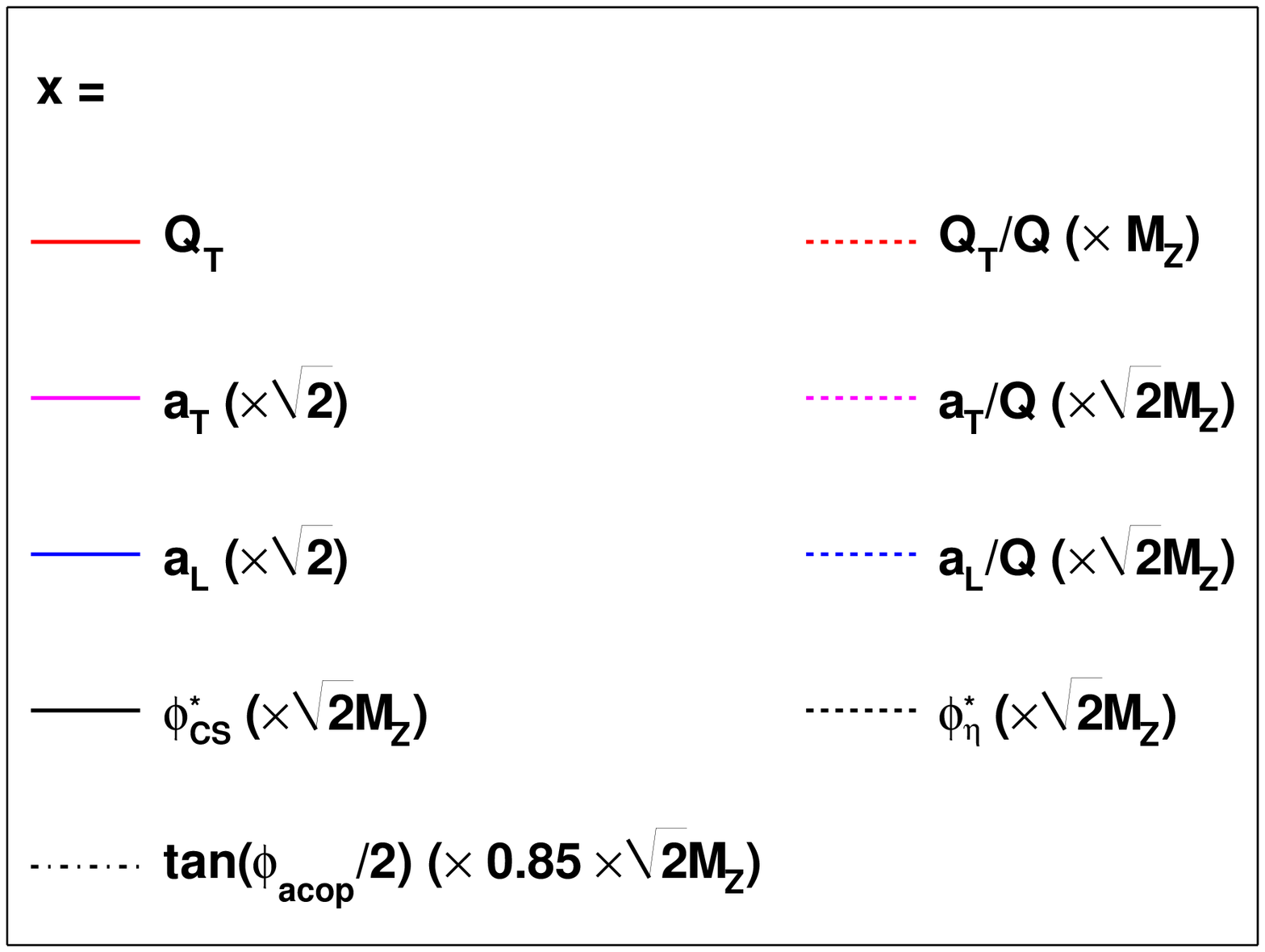}}
    \caption{The mean resolution of each candidate variable,
      as a function of that variable (scaled by the factors described in the text).
      Results are presented both for (a)
      tracker-like and (b) calorimeter-like resolution in the lepton momenta.}
    \label{Figure:resolution_comparison}
  \end{figure}

  Figure~\ref{Figure:resolution_comparison} compares separately for our dimuon and dielectron
  simulations, the mean resolution,
  $|\delta x|$, of various candidate variables, $x$, as a function of (particle level) $x$.
  All variables are scaled by the factors in Table~\ref{Table:ScalingFactors}.
  
  The following observations are made:
  \begin{itemize}
  \item \atom\ is significantly better measured than \at, over the entire range.
  \item Similarly, \Qtom\ is significantly better measured than \Qt.
  \item Over the full range, \at\ and \atom\ perform better than \Qt\ and \Qtom\ respectively.
  \item Compared to \atom, $\phi^{*}_{\rm CS}$ has significantly better resolution,
    and $\phi^{*}_{\rm \eta}$ better still.
  \item The most precisely measured variable is \tanphiaco, since it is determined only from 
    the azimuthal angles of the leptons, whereas the uncertainty on
    \phistarEta\ includes also the uncertainties on the measured
    lepton pseudorapidities.
  \end{itemize}
  
\begin{figure}[htbp]
    \centering
    \subfigure[Tracker]{\includegraphics[width=0.48\textwidth]{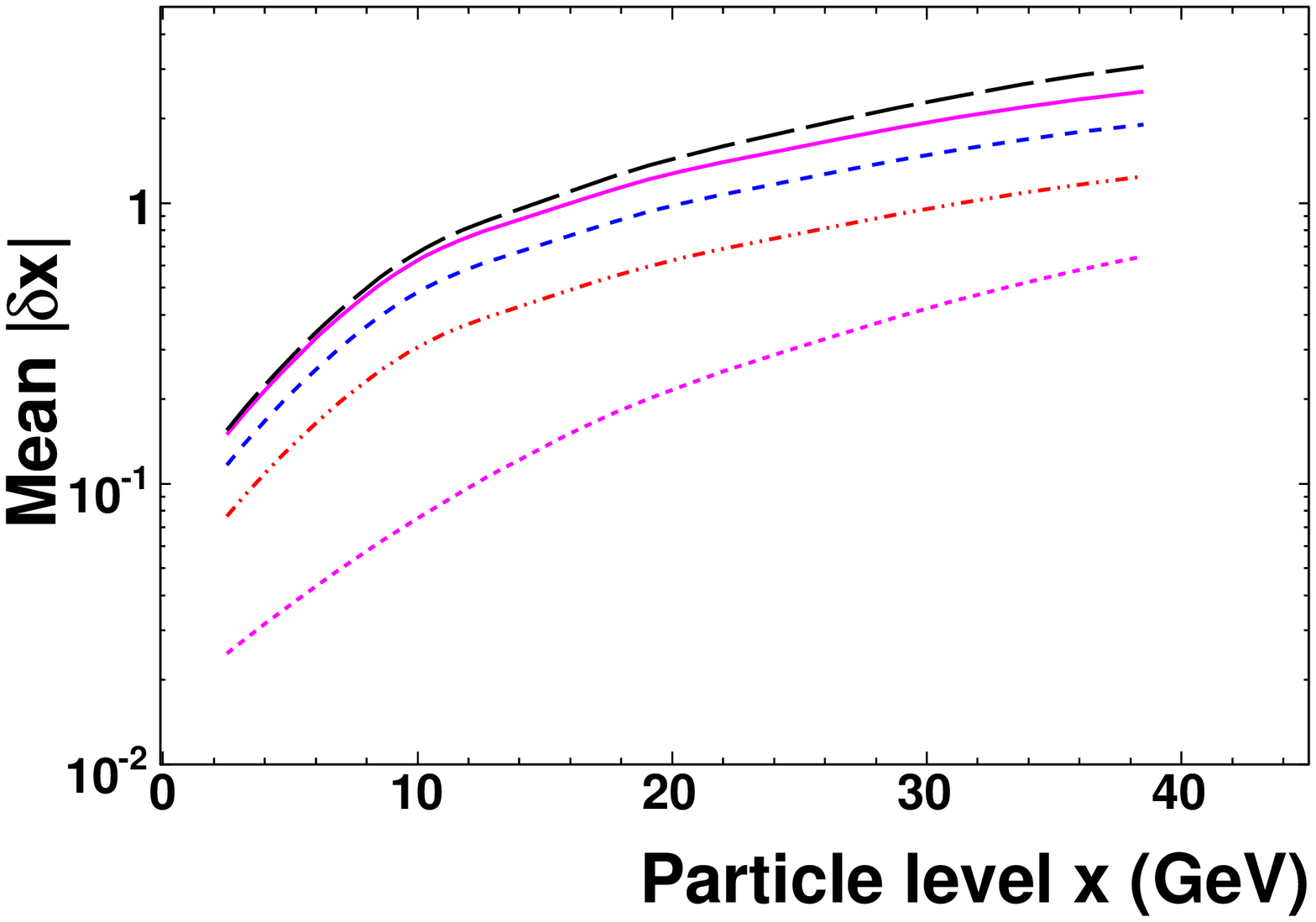}}\\
    \subfigure[Calorimeter]{\includegraphics[width=0.48\textwidth]{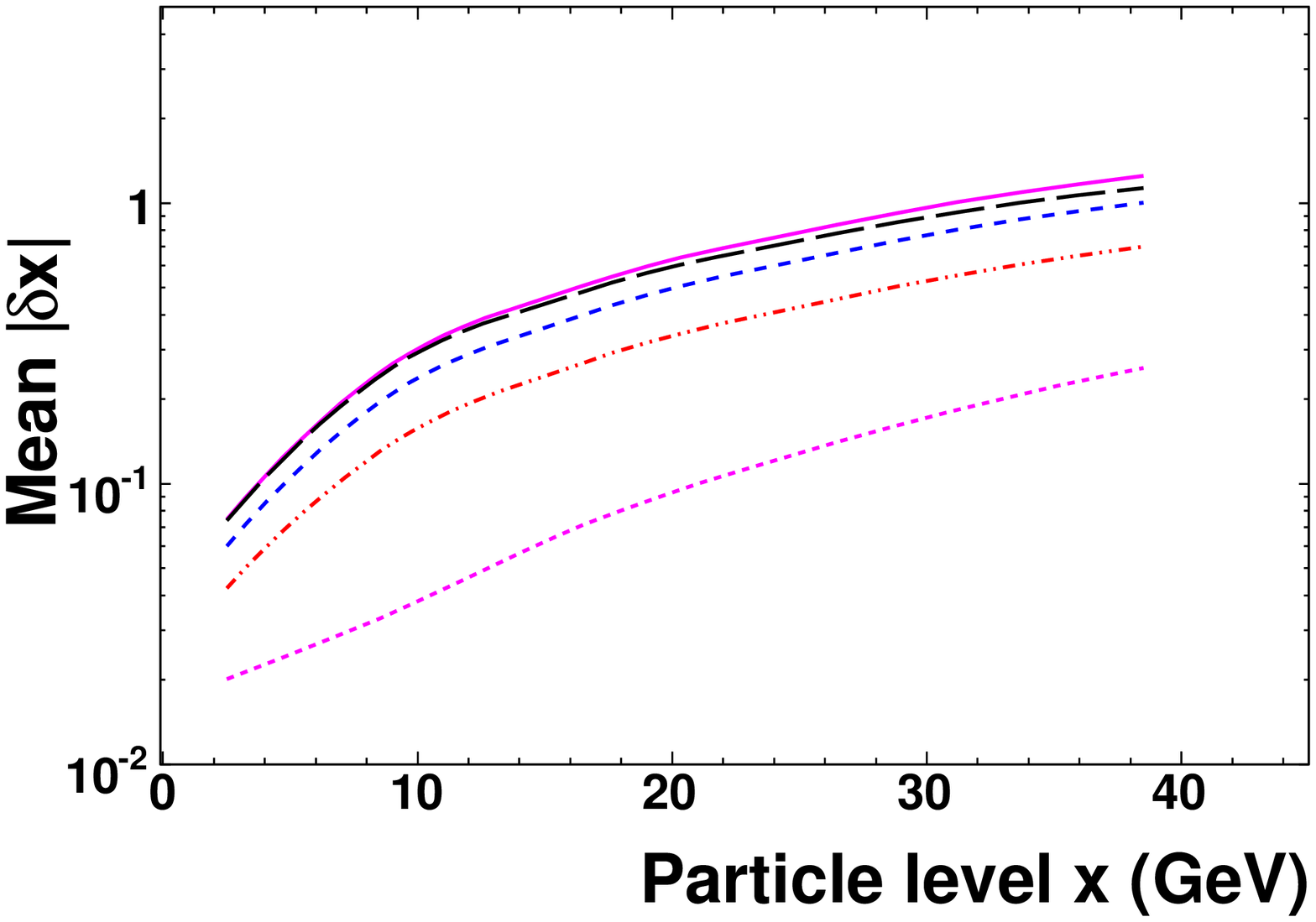}}
    {\includegraphics[width=0.45\textwidth]{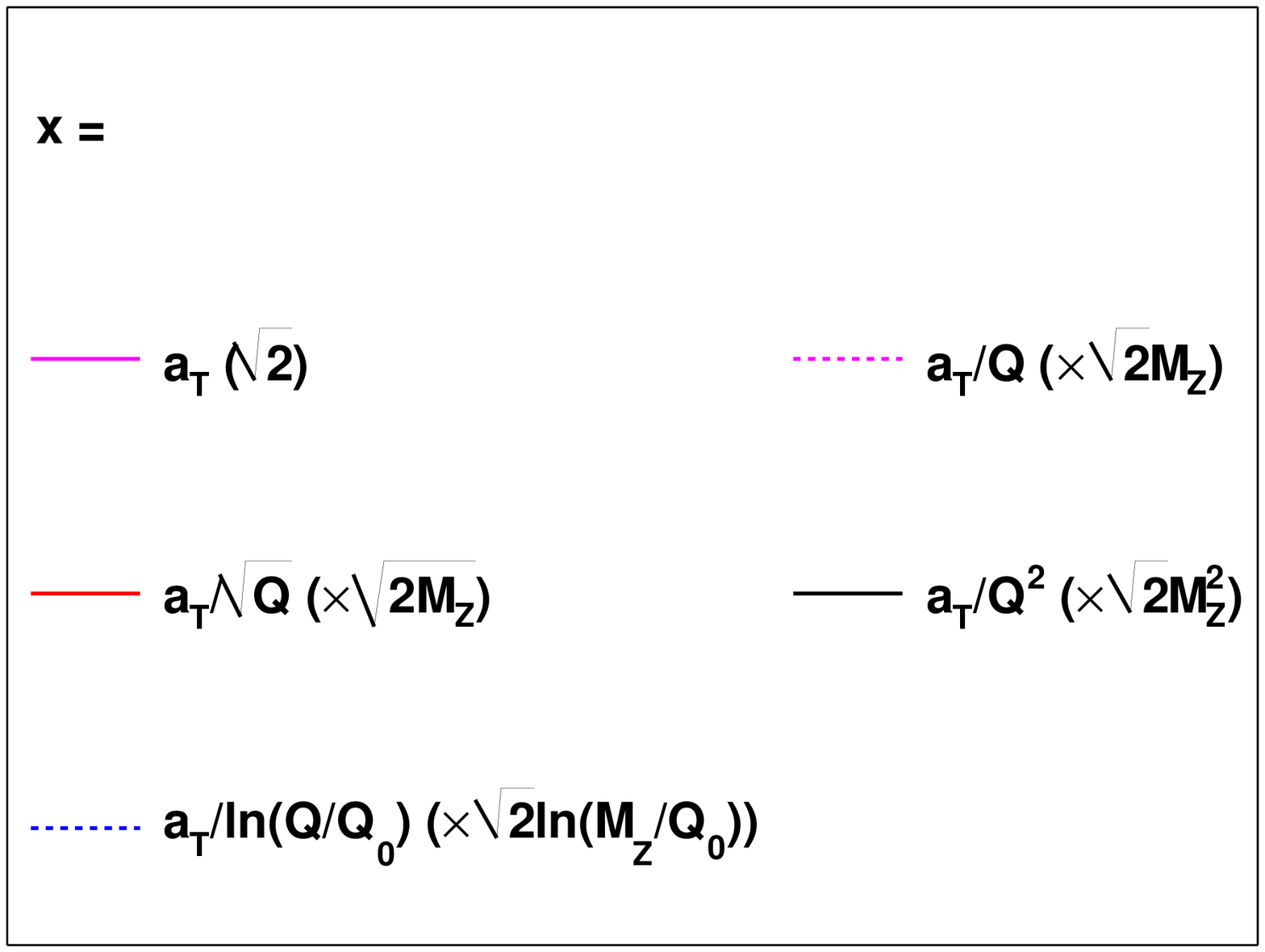}}
    \caption{The mean resolution of the variables \at, \atom, $a_T/\mass^{1/2}$, \atolnm,
      and $a_T/\mass^2$, as a function of that variable (scaled by the factors described in the text).
      Results are presented both for (a)
      tracker-like and (b) calorimeter-like resolution in the lepton momenta.}
    \label{Figure:alternative_comparison}
  \end{figure}

  Since the discussion in Section~\ref{sec-massratio} is only approximate,
  we have investigated empirically various other possible scalings of \at\ with $\mass$
  (with the appropriate scale factor applied, as above).
  These are illustrated in Figure~\ref{Figure:alternative_comparison}.
  As expected, it can be seen that when compared to the other
  variables considered in Figure~\ref{Figure:alternative_comparison}, \atom\ has the best
  experimental resolution for all \Qt\ and irrespective of whether (a)
  tracker-like or (b) calorimeter-like resolution in the lepton
  momenta is simulated.

  \section{Event selection efficiency}
  
  As discussed in~\cite{Mika_Terry_NIM}, the efficiencies of selection cuts on lepton isolation and $p_T$ 
  (for $\at < p_{T}^{\rm cut}$) are less correlated with  \at\ than \Qt.
  We have verified that the correlations of these selection efficiencies
  with \atom\ and \Qtom\ are essentially identical to those with
  \at\ and \Qt\ respectively.
  \dphi\ is primarily sensitive to the \at\ component of the \Qt.
  The variables, \phistarEta, and \phistarCS, 
  are verified to exhibit the same benefits as \at\ compared to \Qt,
  in terms of efficiency dependence.
  
  
  \section{Sensitivity to the physics}
\label{sensitivity}
  
\begin{figure}[htbp]
    \begin{center}
    \includegraphics[width=0.45\textwidth]{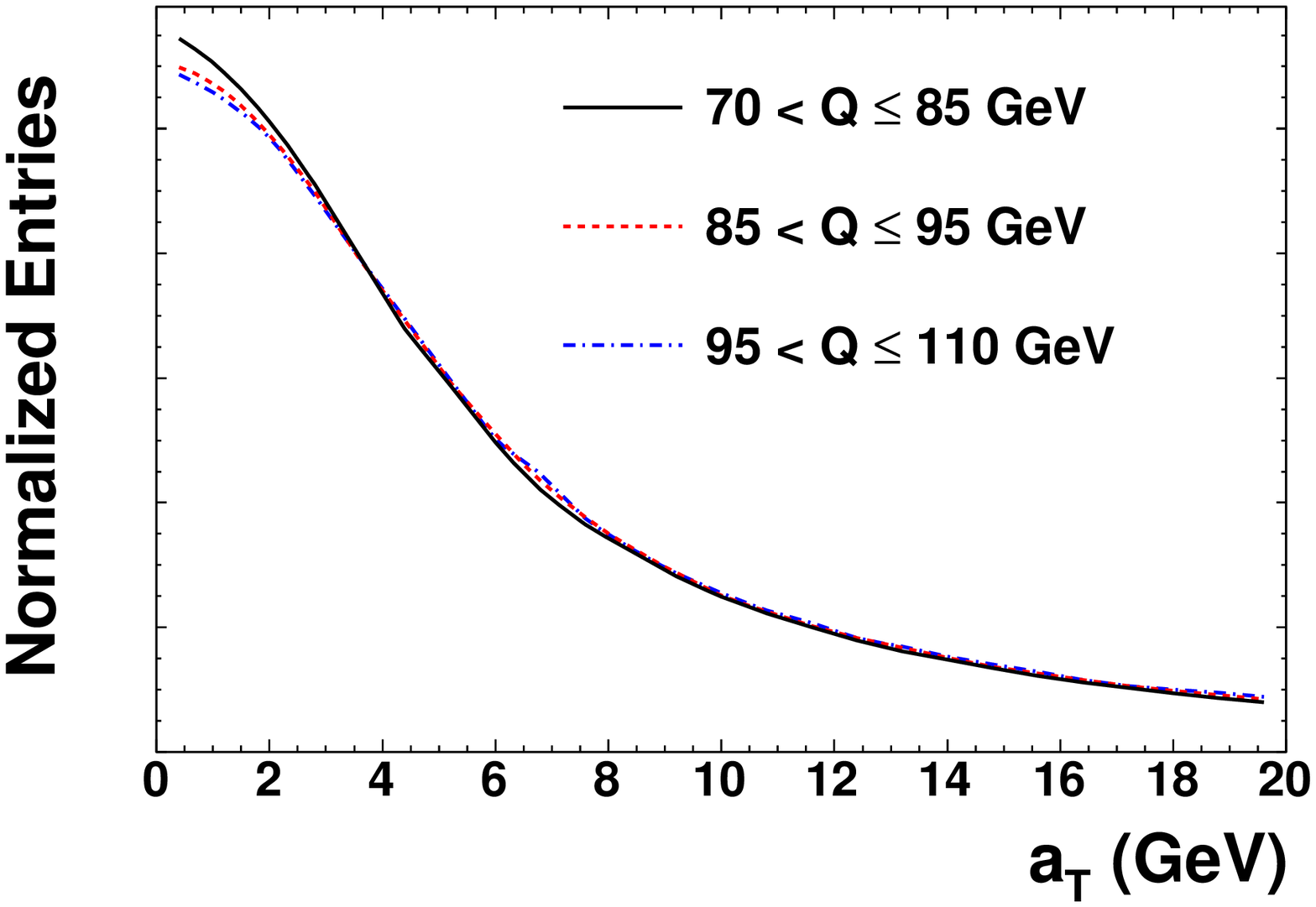}
    \includegraphics[width=0.45\textwidth]{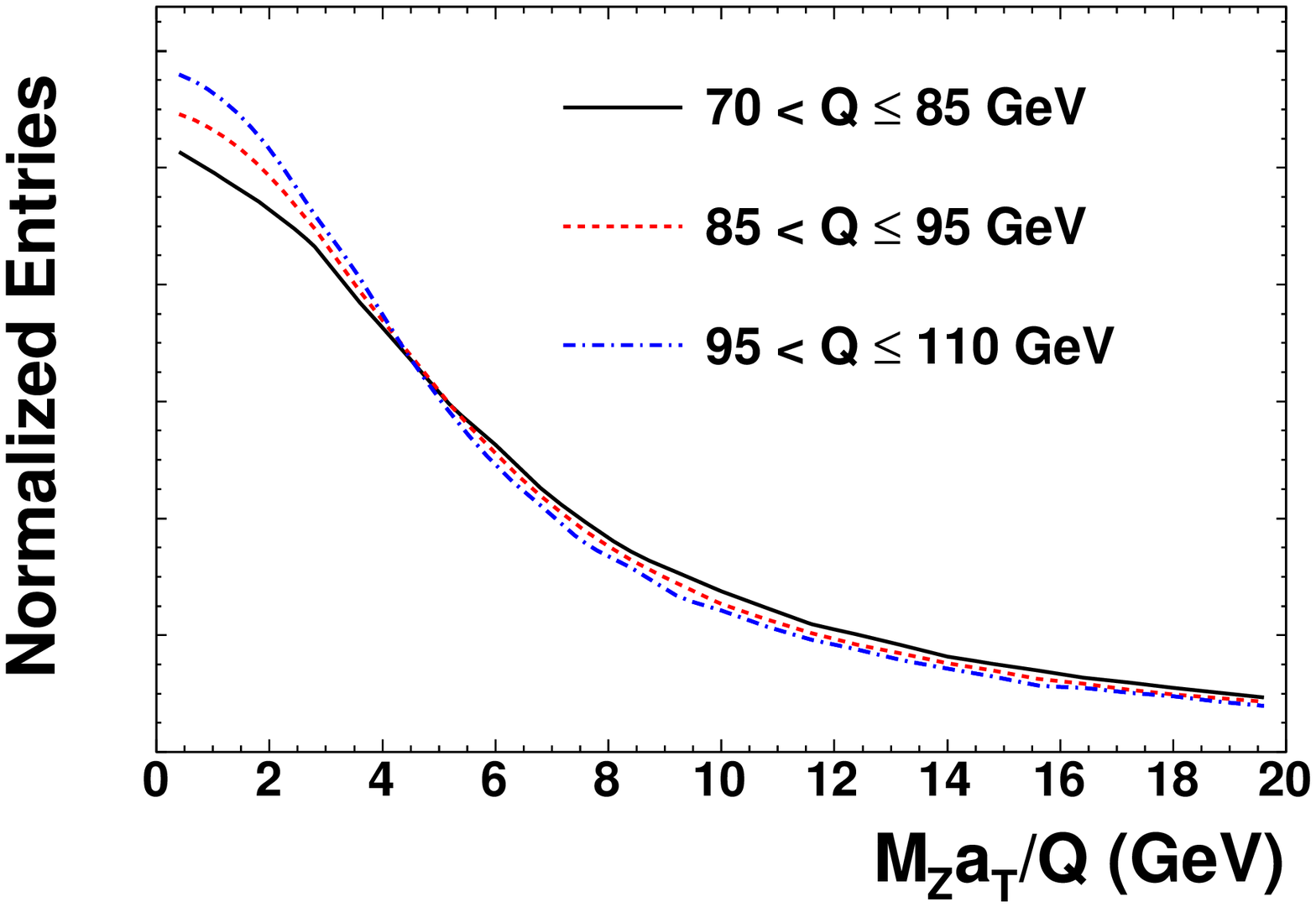}
    \includegraphics[width=0.45\textwidth]{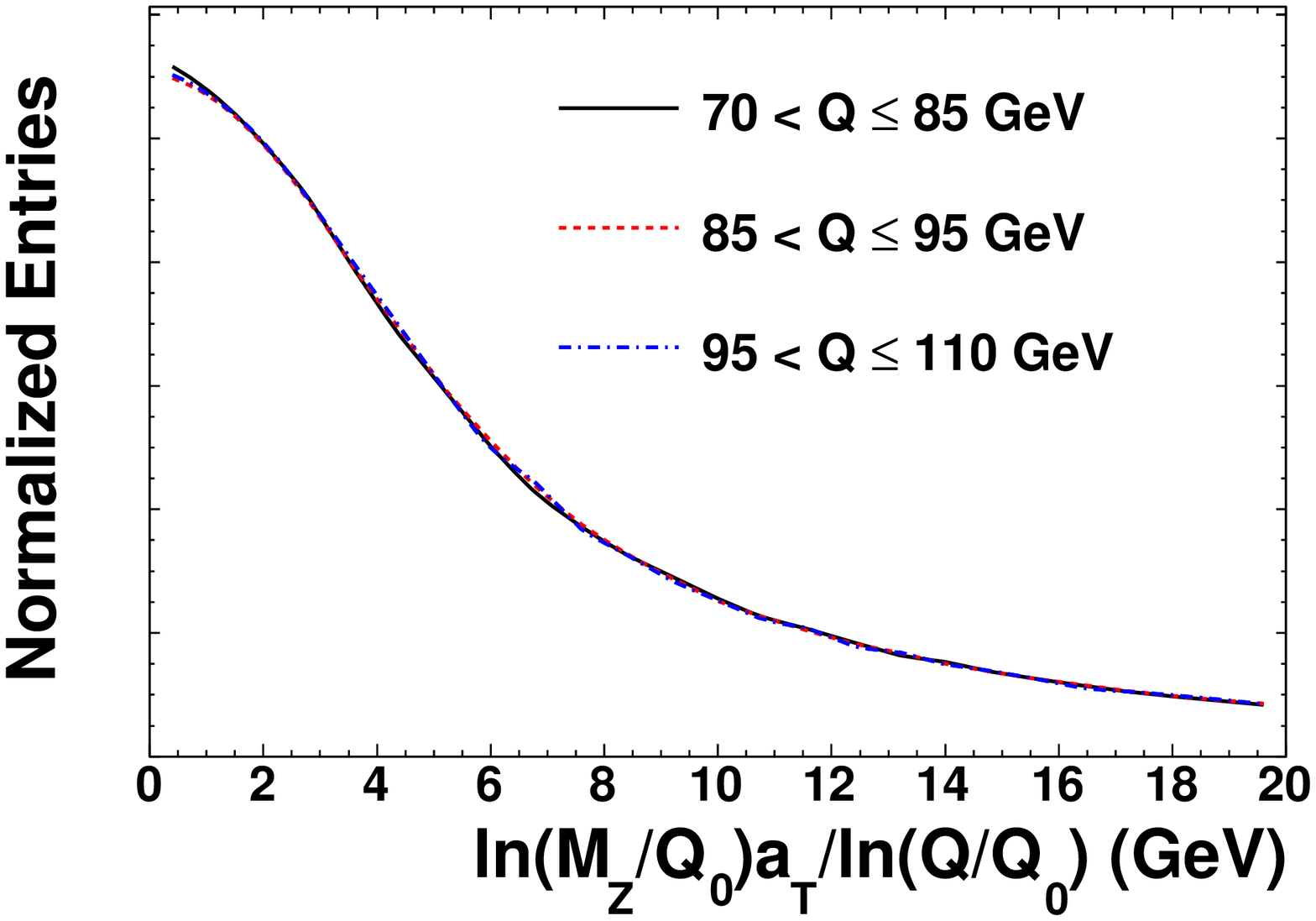}
    \end{center}
    \caption{Comparison of particle level distributions of \at, \matom\ and \matolnm\ for three ranges of \mass.}
    \label{Figure:mass_dependence}
  \end{figure}

Figure~\ref{Figure:mass_dependence} shows the particle level, normalised 
  distributions of \at, \matom\ and \matolnm\ (with $Q_0=1$~GeV) for three ranges of \mass.
We see that \at\ has a mild dependence on $Q$, while dividing by
$Q$ over corrects this dependence. In this respect, we observe that
the distribution in \atolnm\ has a smallest
dependence on $Q$, as might be expected from~\cite{CSS1985}. 

Experimental measurements of $Z$ boson production are typically made
over a fairly wide bin in \mass\ 
  (e.g., 70--110 GeV).  
One potential concern with measurements of \atom\ and \Qtom\
  is that the increased correlation with \mass\ demonstrated in
  Figure~\ref{Figure:mass_dependence} might degrade the sensitivity to the underlying physics.
  Since \phistarEta\ behaves approximately as $\phistarEta \approx \atom$,
  a similar degradation in the physics sensitivity of
  \phistarEta\  may similarly
  be expected.
In this respect, \atolnm\ is a more suitable variable than \atom\ for
studying the boson \Qt\ distribution over a wide range in \mass.
  However it has poorer experimental resolution, as was demonstrated
  in Figure~\ref{Figure:alternative_comparison}.
  
  Using a similar procedure to that described in~\cite{Mika_Terry_NIM},
  we study the sensitivity to the underlying physics of the different
  candidate variables.
  We run pseudo-experiments to fit for the value of the parameter \gtwo,
  which determines the width of the low \Qt\ region in {\sc resbos}.
  Events must meet the requirements described in Section~\ref{fast_sim}, except that these are applied at detector level rather
  than particle level.
  Of these events, 1M are assigned as pseudo-data and the remaining events are used to build
  \gtwo\ templates.
  All variables are scaled by the factors listed in Table~\ref{Table:ScalingFactors},
  such that the same binning (30 equal width bins in the range 0--30 GeV) can be used.
  A minimum \chisqr\ fit
  determines the statistical sensitivity of each variable
  to the value of \gtwo.
  
  Whilst in~\cite{Mika_Terry_NIM} we were mostly interested in the low \Qt\
  region, the ideas proposed in this paper also improve experimental resolution at higher \Qt\ (see Section~\ref{sec-resolution}).
  Thus it is of interest to study the physics sensitivity also in this region.
  We similarly fit for a parameter $K_{Q_T}$, which weights events with (particle level) $\Qt > 25$~GeV,
  by $K_{Q_T}(\Qt - 25)$, and approximately represents the differences between predictions at NLO and NNLO
  discussed in~\cite{DzeroRunII}.
  Again, after applying the appropriate scaling factors from Table~\ref{Table:ScalingFactors}, 
  the same binning can be used for each variable\footnote{
    The first bin is of width 5 GeV (with lower edge at zero) and each consecutive bin is 2 GeV wider than the last.
    Ten such bins give an upper edge of the last bin at 140 GeV and the fit includes the overflow bin from
    140 GeV to $\infty$.
  }.
    
  The results of the fits to \gtwo\ and  $K_{Q_T}$ are presented in
  Tables~\ref{Table:StatErr}~and~\ref{Table:StatErrKF} respectively.
Results are given 
  separately for particle-level (dimuon) and detector-level
  (tracker and calorimeter).
  For both \atom\ and \Qtom, the statistical sensitivities
  are essentially the same as those for \at\ and \Qt\ respectively.
  Thus the effect of the additional \mass\ dependence is shown to be negligible.
  


  \begin{table}[ht]
    \centering
    
    \begin{tabular}{c c c c }
      \hline\hline  
      variable & particle level & calorimeter & tracker \\
      \hline
      \Qt         &  0.65   & 0.94   &  1.41\\
      \Qtom       &  0.65   & 0.94   &  1.40\\
      \at         &  1.00   & 1.00   &  1.00\\
      \atom       &  1.00   & 1.01   &  1.00\\
      \al         &  1.21   & 2.35   &  4.74\\
      \tanphiaco  &  1.04   & 1.05   &  1.04\\
      \phistarCS  &  1.00   & 1.00   &  0.99\\
      \phistarEta &  1.00   & 1.00   &  0.99\\
      \hline\hline

    \end{tabular}
    \caption{Statistical sensitivity (in \%) on the parameter \gtwo\ from fits to the
      distributions of different of variables.  For details see text.}
    \label{Table:StatErr}
  \end{table}

  \begin{table}[ht]
    \centering
    
    \begin{tabular}{c c c c }
      \hline\hline  
      variable & particle level & calorimeter & tracker \\
      \hline      
      \Qt         &  1.65   & 1.67  &  1.82\\
      \Qtom       &  1.66   & 1.67  &  1.78\\
      \at         &  1.92   & 1.92  &  1.96\\
      \atom       &  1.92   & 1.92  &  1.94\\
      \al         &  1.98   & 2.02  &  2.34\\
      \tanphiaco  &  1.96   & 1.96  &  1.98\\
      \phistarCS  &  1.88   & 1.88  &  1.90\\
      \phistarEta &  1.87   & 1.87  &  1.92\\
      \hline\hline

    \end{tabular}
    \caption{Statistical uncertainty (in \%) on the parameter $K_{Q_T}$ (as defined in the text) from fits to the
      distributions of different of variables.  For details see text.}
    \label{Table:StatErrKF}
  \end{table}
  
  The approximately 5\% poorer sensitivity of \tanphiaco, compared to \atom, 
  demonstrates the \sts\ ambiguity of the former.
  The additional factor \sts, present in \phistarEta\ and \phistarCS\ actually recovers the sensitivity 
  to the same level as \at.
  In addition, the \phistarEta\ variable, 
  which was shown in Section~\ref{sec-resolution} to have the best experimental resolution (except for \tanphiaco),
  performs similarly to \phistarCS\ in terms of physics sensitivity.
  
  Of course, the results presented in
  Tables~\ref{Table:StatErr}~and~\ref{Table:StatErrKF} represent only
  the statistical sensitivity of the considered variables when 
  compared to \Qt.
  As discussed in~\cite{Mika_Terry_NIM}, the systematic uncertainties associated with modelling of detector resolution
  and efficiency will be significantly smaller using variables that are better measured and less correlated with 
  event selection efficiencies than is the case for \Qt.


  \section{Summary and conclusions}
 \label{conclusions}
 
  Measurements of dilepton ($Z/\gamma^*$) transverse momentum, \Qt, distributions 
  are crucial for improving models of vector boson production at hadron colliders.
  The precision of future measurements of the \Qt\ distribution at
  the Tevatron and LHC will be 
  totally limited by uncertainties in correcting for detector resolution and efficiency,
  and the minimum bin sizes will be limited by resolution.
  
  In~\cite{Mika_Terry_NIM} an alternative variable, \at, was demonstrated
  to be significantly less susceptible to such detector effects than \Qt.
  We have shown in this article that the experimental resolution of \at\ can be further improved 
  by taking the ratio to the measured dilepton invariant mass, \mass.
  Similarly, we have demonstrated that the variable \Qtom\ is
  experimentally more precisely measured than \Qt.
  No obvious disadvantages of the variable \atom\ (\Qtom) as
  compared to \at\ (\Qt) are found when studying the efficiency
  dependencies or the physics sensitivity (by making fits to parameters describing the shape of the \Qt\ distribution).
  
  The acoplanarity angle, \phiaco, is also sensitive to \Qt.
  However, it has the approximate dependence $\at/\mass \approx \tanphiaco\sin(\theta^*)$,
  where $\theta^*$ is the scattering angle of the leptons with respect to the beam direction
  in the dilepton rest frame. 
  Thus \phiaco\ is less directly related to \Qt\ than \at.
  We show that correcting \tanphiaco\ by factor \sts\
yields a variable with essentially  the same statistical sensitivity as  \atom,
  whilst the experimental resolution is significantly better.
  Furthermore, using an approximate rest frame determined using only
  angular information, we propose a new method to measure the
  scattering angle, $\theta^{*}_{\eta}$, with the best possible
  experimental resolution.
  We conclude that in the region of low \Qt\ the variable
$\phistarEta \equiv \tanphiaco\sin(\theta^{*}_{\eta})$
represents the optimal combination of physics sensitivity,
experimental resolution and
immunity to experimental systematic uncertainties.

  For studying the high \Qt\ region, the optimal variable is \Qtom, 
  which is significantly better measured than \Qt, 
  and has no disadvantages in terms of physics sensitivity or
  efficiency dependence.
However, it is interesting to study correlations between the \at\ and
\al\ components of \Qt.
Therefore, measurements of \phistarEta\ in the high
\Qt\ region will be complementary to measurements of \Qtom. 
  
  These further optimisations of variables used to study the \Qt\ distribution
  will allow significantly finer binning and smaller unfolding corrections (and thus uncertainties),
 enabling tighter constraints on vector boson production models.

\end{document}